\documentstyle[prb,aps,epsf]{revtex}
\begin{document}
\draft
\title{The effect of valley-spin degeneracy on the screening of\\
 charged impurity centers in two- and three-dimensional
 electronic devices}
\author{C. Bulutay}
\address{Department of Electrical and Electronics Engineering, \\
 Middle East Technical University, Ankara 06531, Turkey}
\author{I. Al-Hayek and M. Tomak}
\address{Department of Physics,
 Middle East Technical University, Ankara 06531, Turkey}
\date{\today}
\maketitle
\begin{abstract}
Accurate characterization of charged impurity centers is of
 importance for the electronic devices and materials. The
 role of valley-spin degeneracy on the
 screening of an attractive ion by the mobile carriers is
 assessed within a range of systems from spin-polarized
 single-valley to six-valley. The screening
  is treated using the self-consistent
 local-field correction of Singwi and co-workers known as STLS.
 The bound electron wave function is formulated in the form of an
 integral equation. Friedel oscillations are seen to be
 influential especially in two-dimensions which cannot be
 adequately accounted for by the hydrogenic variational approaches.
 Our results show appreciable differences at certain densities
 with
 respect to simplified techniques, resulting mainly in the
 enhancement
 of the impurity
 binding energies. The calculated Mott constants are provided
 where
 available. The main conclusion of the paper is the substantial
 dependence of the charged impurity binding energy
 on the valley-spin degeneracy in the presence of screening.
\end{abstract}


\section{Introduction}
Impurities are introduced to real electronic systems either
 intentionally as in the form of doping or unintentionally mainly
 in the growth process. Ionization of these impurities contribute
 charged impurity centers to the system and these play important
 role in device operation and material properties; for reviews,
 see Bassani {\it et al.}\cite{bassani} and
 Ando {\it et al.}\cite{ando} for the three-dimensional (3D) and
 two-dimensional (2D) systems respectively. Due to its fundamental
 and technological importance, the binding of electrons to charged
 impurity centers {\it in the presence of screening} has long been
 investigated by the researchers both in
 3D\cite{k&n,greene,aldrich,neet83,martino,neet85,gold96} and
 2D\cite{panat,brum,degani,andrada}.
 Most of these works in 3D aimed to deal with the metal-insulator
 transition using\cite{kamim,mott90} Mott's initial approach which
 he devised, in fact, for the stability of the metallic phase.
 These approaches can be grouped as variational
\cite{k&n,greene,aldrich,neet83,panat,brum,degani,andrada}
 and numerical\cite{martino,neet85,gold96} treatment of the
 bound electron wave function. The former has appealing simplicity
 but may not be suitable to use in this problem as the resultant
 binding energy will inevitably be higher than the true ground-state
 energy. Martino {\it et al.}\cite{martino} have shown this to
 be the case by comparing Krieger and Nightingale's\cite{k&n}
 variational hydrogenic wave function treatment with their numerical
 method. Consequently, variational techniques are refined by
 replacing the trial wave functions with the Hulth\'en's
 form,\cite{hulthen} resulting in a better agreement with
 Ref.\ \onlinecite{martino}. In all these works
 several forms of dielectric screening have been employed such as
 Thomas-Fermi,\cite{k&n,greene,aldrich} Random Phase Approximation
 (RPA),\cite{k&n,greene,aldrich,neet83} 
 and Hubbard-Sham.\cite{martino,greene,neet85}
 In our work we use the so-called self-consistent local-field
 correction of Singwi and co-workers\cite{stls} to be referred to
 as STLS. This technique is one of the best improvements of the
 RPA,\cite{mahan} considering both the Pauli and Coulomb holes
 around the electrons that take part in the screening of any
 longitudinal disturbance.\cite{mahan} Very recently Borges
 {\it et al.}\cite{borges} also dealt with impurity binding
 energies in 3D using STLS dielectric function, with Hulth\'en
 variational wave function. 

 The aim of this work is primarily to assess the role of valley
 and spin degeneracy on the screening and hence binding strength
 of charged impurity centers both in 2D and 3D. The valley and
 spin degeneracy means, a number of conduction band minima and
 two spin states are {\it energetically} degenerate in the
 absence of symmetry-breaking perturbations like strain and
 magnetic field. The need for this work stems from the current
 status of electronic devices, evolved into two classes as:
 multi-valley systems dominated by silicon and germanium and
 single-valley systems realized using mainly gallium arsenide.
 Another source of motivation for the present work is
 related to the recently observed Metal-Insulator transition in
 a 2D system, Si MOSFET, towards zero
 temperature.\cite{krav96,furn96,krav95,krav94} We
 commented\cite{bul_com} that this observation could be due to
 valley phase transition in Si inversion layers, a many-body effect
 originally anticipated by Bloss, Sham and Vinter.\cite{bloss}
 For this reason, the effect of valley degeneracy on several
 physical phenomena needs further investigation; the screening of
 charged impurity centers is just one of them.
 We observe that Friedel oscillations\cite{mahan,fetter} associated
 with the screening of an impurity potential are influential in the
 impurity binding energies. Appreciable differences are
 seen between our numerical approach based on the
 computationally-efficient solution of an integral equation and
 the widely-used hydrogenic variational treatments. These results
 will have implications on the accurate characterization of
 impurities as well as excitons\cite{haug} under the presence of
 free carrier screening. The screened attractive impurity potential
 has two important aspects: the bound-states and the scattering
 cross section. In this paper we focus on the former and do not
 address the equally-important problem of the effect of
 valley-spin degeneracy on the {\it mobility}, limited by the
 screened ionized impurities.

In Sec.~II the theoretical details on the variational expressions,
 the dielectric screening and the integral equation formulation
 are included. The results for 3D and 2D cases can be found in
 Sec.~III. We observe that the phenomena in 2D is quite interesting
 and thereby requires more discussion. Finally, we conclude in
 Sec.~IV. Throughout the text we mention the assumptions and
 simplifications made.

\section{Theoretical Details}
 We assume that the mobile carriers are over a neutralizing
 positively charged continuum, that is,
 the so-called electron liquid (EL) model, where electron-electron
 interactions are rigorously considered using the dielectric
 formulation\cite{pn} but ionic lattice and disorder effects are
 ignored. We introduce a degeneracy parameter for the constituent
 electrons of the EL, and thereby consider the spectrum ranging
 from spin-polarized and single-valley EL to
 six-valley EL to assess these exchange effects due to valley-spin
 degeneracy. We consider one singly-ionized impurity\cite{chemical}
 being immersed into an EL, and investigate whether this impurity
 can trap an electron with the screening of the EL present. 
 The binding ability in this model is controlled by three effects:
 i) the attractive bare Coulomb interaction
 of the ion that enhances the binding, ii) the kinetic energy of
 the bound electron that tries to overcome the binding and iii)
 the screening of the bare interaction by the free carriers of the
 EL that weakens the binding of the electron. A negative energy
 bound-state may not be possible if the last two effects win over
 the first as the concentration of the free carriers in the EL is
 increased. We observe that the competition among these effects
 becomes more interesting in the 2D case. The zero intercept
 density of the binding energy is traditionally known as the
 Mott value, which renders the comparison of different approaches
 quite simple; we display our results for the Mott constant in
 Sec.~III. Our single impurity consideration poses one important
 restriction on the density of impurity centers; basically, if
 the bound electron wave functions overlap due to large impurity
 concentration, then the discrete bound-states broaden into bands
 by means of tunnelling between neighbouring sites. However, the
 aim in high speed electronic devices is to avoid large amount of
 impurities along the mobile carrier paths.

We work at zero temperature, and for the 2D case, aiming for
 general results, we assume no extension along the third
 dimension (i.e., strictly 2D), where electrons still interact
 with Coulomb $1/R$ potential.\cite{isihara} We mainly use 3D
 effective Rydbergs ($\mbox{Ry}^*$) for the energies and denote
 them with an overbar. Also we introduce length-related reduced
 variables by scaling with the effective Bohr radius, $a^{*}_{B}$
 and denote them by the subscript $r$ throughout the text such as
 $a_r\equiv a/a^{*}_{B}$, $q_r\equiv a^{*}_{B}q$, for variables
 having length and reciprocal length dimensions respectively.  

\subsection{Variational Expressions}
For completeness we first list the expressions for the hydrogenic
 variational approach. In 3D case Krieger and Nightingale\cite{k&n}
 used a variational approach for the bound electron wave function
 based on a hydrogenic $1s$ trial wave function as 
\begin{equation}
\label{3dwf}
\psi_0(r)=\frac{1}{\sqrt{\pi a^3}}\: e^{-r/a},
\end{equation}
 with $a$ being the variational parameter, whereas Panat
 and Paranjape\cite{panat} used the same form in 2D with radial
 variable $r$ being replaced by the polar radial variable $\rho$ as
\begin{equation}
\label{2dwf}
\psi_0(\rho)=\sqrt{\frac{2}{\pi a^2}} \: e^{-\rho /a}.
\end{equation}
The corresponding binding energies are given as
\begin{equation}
\overline{E}_0(a_r)=\frac{1}{a_r^2}-\frac{4}{\pi}\left\{\int_0^\infty
\frac{dq_r}{\left [ \left ( \frac{a_r q_r}{2}\right )^2 +1\right ]^2 }
\left [ \frac{1}{\epsilon_{3D}(q_r)}-1\right] \right\} -\frac{2}{a_r},
\end{equation}
in 3D and
\begin{equation}
\overline{E}_0(a_r)=\frac{1}{a_r^2}-2\left\{\int_0^\infty
\frac{dq_r}{\left [ \left ( \frac{a_r q_r}{2}\right )^2 +1\right ]^{3/2}
 }\left [ \frac{1}{\epsilon_{2D}(q_r)}-1\right] \right\} -\frac{4}{a_r},
\end{equation}
in 2D case, where bare interaction is added and subtracted to achieve
 faster decaying integrand as suggested in Ref.\ \onlinecite{panat}.
 In these expressions $\epsilon_{3D}$ and $\epsilon_{2D}$ represent
 the static dielectric screening of the EL, which can be computed at
 different levels of sophistication; in the next section we describe the
 dielectric function we employ.

\subsection{Dielectric Function For A General Degeneracy Factor}
\begin{figure}[t]
\vspace*{-4cm}
\centerline{
\epsfxsize=12cm \epsfbox{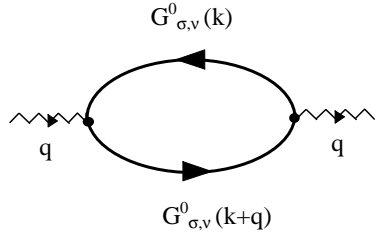}
}
\vspace*{-6cm}
\caption{Zeroth-order polarization insertion diagram, 
  $\pi^0(q)$ for a fermionic system having spin 
 ($\sigma$), valley ($\nu$), and wave number ($k$) labels.}
\label{Fpi0}
\end{figure}

The ultimately important quantity in screening is the static
 dielectric function of the EL. Here we would like to introduce an extra
 label ($\nu$), to the free
 carriers of the EL in addition to spin $(\sigma)$ and wave vector
 ($\vec{k}$) labels, to account for the extra valley freedom\cite{note2}.
 Then, the zeroth-order polarization insertion diagram\cite{fetter}
 $\mbox{\large $\pi^0$}(q)$  
 is modified as in Fig.~\ref{Fpi0}, where $G^0$ refers to noninteracting
 propagator.\cite{fetter} The bare Coulomb interaction being independent
 of spin and valley labels, suggests the introduction of an overall
 degeneracy factor $g_d$ as $g_d=g_s g_v$ where $g_s$ and $g_v$ are the
 spin and valley degeneracies respectively. For the spin-polarized
 single-valley EL we have $g_d=1$ and for the normal-state EL having
 $g_v$ energetically degenerate valleys, $g_d=2g_v$.
 In turn, $\mbox{\large $\pi^0$}$ is trivially affected by the overall
 degeneracy factor
 $g_d$ as a coefficient in front. The 3D static dielectric function
 becomes
\begin{equation}
\epsilon_{3D}^{STLS}(q_n)=\frac{1+\left(\frac{2g_d^4}{9\pi^4}\right)^{1/3}
\frac{r_s}{q_n^2}\left(1-\frac{4-q_n^2}{4q_n}\ln\vert\frac{2-q_n}{2+q_n}
\vert
\right)\left[1-G_{3D}^{STLS}(q_n)\right]}{1-\left(\frac{2g_d^4}{9\pi^4}
\right)^{1/3}\frac{r_s}{q_n^2}\left(1-\frac{4-q_n^2}{4q_n}\ln\vert\frac
{2-q_n}{2+q_n}\vert\right)G_{3D}^{STLS}(q_n)}.
\end{equation}
In this expression and throughout the text $q_n$ refers to a wave number
 normalized to Fermi wave number $k_F$ and $r_s=1/a_{B}^{*}(\frac{4}{3}\pi
 n_{3D})^{1/3}$, with $n_{3D}$ being 3D free electron density. $q_r$ and
 $q_n$ are related in 3D as $q_n=q_r r_s (2g_d/9\pi)^{1/3}$. The Pauli
 and Coulomb holes surrounding the screening electrons are introduced by
 the local-field correction $G_{3D}^{STLS}$, which needs a self-consistent
 calculation as described in Ref.\ \onlinecite{stls}.
 Recently, Gold\cite{gold94} has
 investigated the effects of valley degeneracy on the local-field
 correction. He noted that the many-body effects are important even at
 high electron densities (i.e., low $r_s$ values). We find Gold's work
 very useful, however, Gold constrained the local-field correction to a
 Hubbard-like form\cite{hubbard57} that led to simplicity in the
 computation.
 Based on our previous observations,\cite{bulutay96b} we avoid this
 simplification
 and use the standard approach. 

The static dielectric function in 2D is of the form
\begin{equation}
\label{eq:*}
\mbox{\Large $\epsilon$}^{STLS}_{2D}(q_n)=\left\{
\begin{array}{ll}
\frac{\displaystyle 1+\frac{g_d^{3/2}r_s}{2q_n}\left[1-G_{2D}^{STLS}
(q_n)\right]}
{\displaystyle 1-\frac{g_d^{3/2}r_s}{2q_n}G_{2D}^{STLS}(q_n)}
 & \mbox{for $q_n\leq2$} \\ \\
\frac{\displaystyle 1+\frac{g_d^{3/2}r_s}{2q_n}\left[1-\sqrt{1-(\frac
{2}{q_n})^{2}}\,\right]\left[1-G_{2D}^{STLS}(q_n)\right]}
{\displaystyle 1-\frac{g_d^{3/2}r_s}{2q_n}\left[1-\sqrt{1-(\frac{2}{q_n}
)^{2}}\,\right]G_{2D}^{STLS}(q_n)}
 & \mbox{for $q_n>2$}
\end{array} \right. ,
\end{equation}
where $r_s$ in this case is related to 2D electronic density, $n_{2D}$
 by $r_s=1/a_B^* \sqrt{\pi n_{2D}}$. The relation between $q_n$ and $q_r$
 in 2D is $q_n=q_r r_s \sqrt{g_d}/2$.
Again the 2D local-field correction, $G_{2D}^{STLS}(q_n)$ needs to be 
 self-consistently determined at each value of $r_s$. We refer to
 available literature\cite{jonson,bulutay96a} for details, however,
 we would like to mention the work of de Freitas
 {\it et al.},\cite{freitas} which greatly simplifies the labour in
 the static structure factor calculation by the so-called
 Ioriatti-Isihara\cite{ioriatti}
 transformations. We applied their recipe to a 2D EL having an arbitrary
 spin and valley degeneracy.

\subsection{Integral Equation Formulation}
In contrast to simplicity of the variational techniques,
 they must be used with care in problems such as the binding energy,
 where the variational energy only yields an upper bound for the true
 ground-state energy. As a better alternative, we present below an
 approach that leads to an integral equation for which we also develop
 computationally-efficient operator techniques.

\subsubsection{Formulation for 3D}
The bound electron feels a centrally symmetric radial potential, and for
 the lowest $1s$ state, Schr\"odinger equation in 3D becomes
\begin{equation}
\label{radial}
\frac{1}{r_r^2}\frac{d}{dr_r}\left(r_r^2\frac
{d\psi_0(r_r)}{dr_r}\right)+\left[\/ \overline{E}_0-\overline{U}_{3D,scr}
(r_r)\right]\psi_0(r_r)=0 \/.
\end{equation}
$\overline{U}_{3D,scr}(r_r)$ is the screened potential energy due to a
 singly-ionized
 attractive impurity in real space to be computed as
\begin{equation}
\overline{U}_{3D,scr}(r_r)=-\frac{2}{r_r}+\frac{4}{\pi}\int_0^\infty dq_n
\frac{\sin \left(q_n\frac{r_r}{r_s}\left[\frac{9\pi}{2 g_d}
\right]^{1/3}\right)}{q_nr_r}\left[1-\frac{1}{\epsilon_{3D} (q_n)}\right],
\end{equation}
where $r_r$ is the reduced distance in real space; again we add and
 subtract unscreened Coulomb potential for computational
 reasons.\cite{panat}

The principal problem in the numerical solution is the infinite domain of
 the wave function. As a remedy, Martino {\it et al.}\cite{martino}
 noting the difference with the hydrogen atom problem due to the presence
 of screening, set $\overline{U}_{3D,scr}(r)$ to zero for distances
 greater than some large value $R$. Then, for
 $r>R$ the wave function for bound-states becomes\cite{neet85}
\begin{equation}
\psi_0(r)\sim \frac{e^{-\kappa r}}{r} \: \mbox{with}\:\kappa=
\sqrt{\frac{2m^*}{\hbar^2}\vert E_0\vert} ,
\end{equation}
upto a normalization constant.  The continuity of the
 wave function together with its derivative at $r=R$, or equivalently,
 the continuity of the logarithmic derivative of the wave function
 yields
\begin{equation}
\left. \frac{d\ln \psi_0(r)}{dr}\right\vert_{r=R}=-\kappa-\frac{1}{R}.
\end{equation}
 Eq.~(\ref{radial}) is a two-value differential equation problem dealt
 with shooting type numerical techniques.\cite{press} We, rather,
 prefer to convert the radial Scr\"odinger equation to an integral
 equation as
\begin{equation}
\label{gamma(r)}
\Gamma(r_r)=S(r_r)-\frac{1}{r_r}\int_{0}^{r_r}dr^\prime\:
 \Gamma(r^\prime)^2 \/,
\end{equation}
where 
\[S(r_r)=\frac{1}{r_r}\int_{0}^{r_r}dr^\prime r^{\prime^2}\:
 \overline{U}_{3D,scr}(r_r^\prime)+\frac{r_r^2}{3}\vert
 \overline{E}_0\vert ,\]
and 
\begin{equation}
\Gamma(r_r)=r_r\frac{d\ln \psi_0(r_r)}{dr_r}.
\end{equation}
 The energy eigenvalue is determined from $\Gamma(R_r)=-R_r\sqrt{\vert
 \overline{E}_0\vert}-1$. In our work we extracted the energy
 eigenvalue $\overline{E}_0$ from the above equation by
 sampling the wave function in the 5\% neighborhood of $R_r=10$.
Eq.~(\ref{gamma(r)}) is a nonlinear integral
 equation of the Volterra type in the fixed-point form.\cite{zeidler}
 However, we observed very
 slow convergence of the standard techniques; for this purpose we
 first express the Eq.~(\ref{gamma(r)}) as an operator equation as
\begin{equation}
P\left[ \Gamma(r_r)\right]=\Gamma(r_r)-S(r_r)+\frac{1}{r_r}\int_0^{r_r}
 dr^\prime\: \Gamma(r^\prime)^2=0
\end{equation}
 and resort to operator form of the Newton's method\cite{rall} which
 requires the inverse operator of the derivative (Fr\'echet
 derivative\cite{rall}) of the operator $P$ evaluated at
 the function $\Gamma(r_r)$. This inverse operator acting on
 $\Gamma(r_r)$ is given as
\begin{equation}
\left\{ P^\prime\left[\Gamma(r_r)\right]\right\}^{-1}=
\sum_{n=0}^\infty (-1)^n \frac{2}{r_r} \int_0^{r_r}
 dr^\prime \Gamma(r^\prime) \frac{2}{r^\prime} \int_0^{r^\prime}
 dr^{\prime\prime} \Gamma(r^{\prime\prime}) \cdots \frac{2}{r^{(n-1)}}
 \int_0^{r^{(n-1)}} dr^{(n)} \Gamma(r^{(n)});
\end{equation}
here the prime on the left hand side designates a derivative, whereas,
 the primes on the right hand side are used to produce dummy variables.
We retain the first two terms and approximate the final equation as
\begin{equation}
\Gamma_{new}(r_r)=\Gamma_n(r_r)-P\left[ \Gamma_n(r_r)\right]+
\frac{2}{r_r}\int_0^{r_r} dr^\prime \: \Gamma_n(r^\prime)P
\left[ \Gamma_n(r^\prime)\right],
\end{equation}
where $\Gamma_n$ denotes  $\Gamma$ at the $n^{th}$ iteration;
 we determine $\Gamma_{n+1}$
 by mixing $\Gamma_{new}$ and $\Gamma_n$. The final form, then
 offers a rapidly converging algorithm, once we initiate the
 process at low densities (like $r_s=20$)
 using the variational wave function as the initial guess and
 gradually increase the density.

\subsubsection{Formulation for 2D}
\begin{figure}[b]
\vspace*{1.5cm}
\centerline{
\epsfxsize=8cm \epsfbox{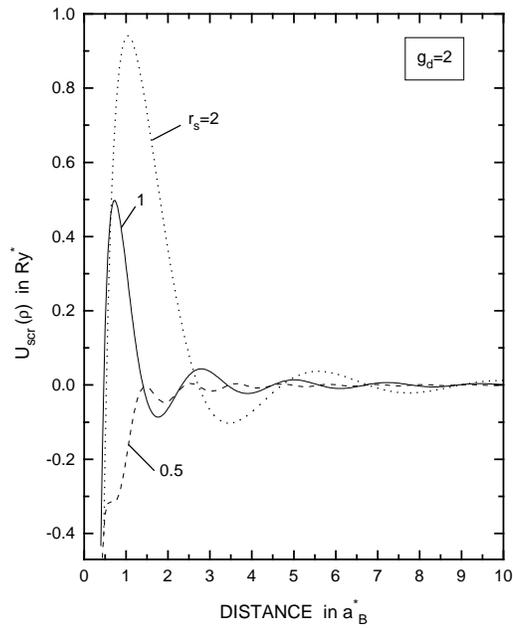}
}
\vspace*{-0cm}
\caption{Potential energy distribution due to a screened,
 singly-ionized attractive impurity versus distance.
 A normal-state, single-valley EL is considered ($g_d=2$)
 at several densities.}
\label{Fuscr1}
\end{figure}

2D Schr\"odinger equation for the ground-state wave function reads
\begin{equation}
\frac{1}{\rho_r}\frac{d}{d\rho_r}\left (\rho_r\frac{d\psi_0(\rho_r )}
{d\rho_r} \right ) + \left ( \overline{E}_0 - \overline{U}_{2D,scr}
(\rho_r)\right ) \psi_0(\rho_r)=0.
\end{equation}
 The expression for the screened potential energy due to a
 singly-ionized attractive impurity in real space, which shows
 Friedel oscillations is
\begin{equation}
\overline{U}_{2D,scr}(\rho_r)=-\frac{2}{\rho_r}+\frac{4}{r_s \sqrt{g_d}}
\int_0^\infty dq_n\:J_0\left(q_n\frac{2}{r_s \sqrt{g_d}}\rho_r\right)
\left[1-\frac{1}{\epsilon_{2D} (q_n)}\right],
\end{equation}
where $\rho_r$ is the reduced 2D radial coordinate and $J_0$
 is the zeroth-order cylindrical Bessel function of the first kind.
 See Fig.\ \ref{Fuscr1} for the screened potential energy at
 several values of the 2D electronic density; also note the evolution
 of the Friedel oscillations as the density decreases.
As in 3D, we work with the function
 $\Gamma(\rho_r)=\rho_r d\ln\psi_0/d\rho_r$,
 rather than with the wave function itself; in this way an exponentially
 decaying function is mapped to a linearly decreasing one. The nonlinear
 integral equation satisfied by $\Gamma$ becomes
\begin{equation}
\Gamma(\rho_r)=S(\rho_r)-\int_0^{\rho_r} d\rho^\prime
 \frac{\Gamma(\rho^\prime)^2}
{\rho^\prime},
\end{equation}
where 
\begin{equation}
S(\rho_r)=\int_0^{\rho_r}d\rho^\prime \rho^\prime\: \overline{U}_{2D,scr}
(\rho^\prime )+\frac{\rho_r^2 |\overline{E}_0|}{2},
\end{equation}
which is to be computed with very high precision. A nonlinear equation
 needs to be solved for the bound-state energy eigenvalue, of the form
\begin{equation}
\Gamma(R_r)=-R_r\sqrt{|E_0|}
K_1(R_r\sqrt{|E_0|})/K_0(R_r\sqrt{|E_0|}),
\end{equation}
where $K$ is the modified Bessel function of the second kind.

To achieve much faster convergence than the fixed-point form,
 the operator $P$ is introduced as
\begin{equation}
P\left[ \Gamma(\rho_r)\right]=\Gamma(\rho_r)-S(\rho_r)+\int_0^{\rho_r}
d\rho^\prime \frac{\Gamma(\rho^\prime)^2}{\rho^\prime}=0
\end{equation}
We give the final form of the iterative equation we use in 2D which
 closely resembles the 3D case
\begin{equation}
\Gamma_{new}(\rho_r)=\Gamma_n(\rho_r)-P\left[ \Gamma_n(\rho_r)\right]+
2\int_0^{\rho_r} d\rho^\prime \frac{\Gamma_n(\rho^\prime)}
{\rho^\prime}P\left[ \Gamma_n(\rho^\prime)
\right].
\end{equation}

\section{Results}
\subsection{3D EL Results}
\begin{figure}[t]
\vspace*{-1cm}
\centerline{
\epsfxsize=7.5cm \epsfbox{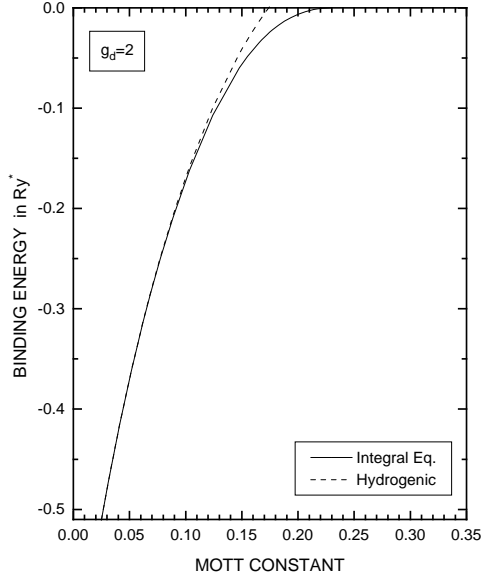}
}
\vspace*{-0cm}
\caption{Binding energy of a bound impurity electron within a
 normal-state, single-valley ($g_d=2$) 3D EL versus the Mott
 constant, defined as $a_{B}^{*}n_{3D}^{1/3}$. Solid line refers
 to integral equation solution which gives a lower energy than the
 variational treatment based on the hydrogenic wave function denoted
 by the dashed lines.}
\label{Fmott3d}
\end{figure}

 We investigate the binding energy of the impurity electron as a function
 of the electron density and valley-spin degeneracy. In Table~\ref{3dmott},
 we list the so-called Mott constant as a function of the degeneracy
 parameter $g_d$ from spin-polarized electrons to six valley degeneracy
 as in the conduction band of silicon. Here Mott constant is defined as
 $a_{B}^{*}n_{3D}^{1/3}$, where $n_{3D}$ is the density at which the
 binding energy reaches zero. It can be seen that for $g_d$ greater
 than 4 the exchange
 effects do not lead to appreciable changes in the Mott constant. The
 spin-polarized EL ($g_d=1$) has the highest Mott constant, which is
 due to poor screening of the impurity potential by the participating
 electrons having large Pauli holes around them.
 In Fig.~\ref{Fmott3d} we plot the variational (hydrogenic)
 and integral equation solutions for the binding energy of the normal-state
 single-valley ($g_d=2$) EL;
 the deviation is clearly visible towards the Mott constant. The
 behaviour for other valley-spin degeneracies is the same apart from a
 translation according to the Mott constant value; refer to
 Table~~\ref{3dmott}.

Our treatment is based on an isotropic effective mass for the screening
 electrons, however, mass anisotropy is predominantly effective in
 multi-valley materials such as silicon and germanium. We refer to available
 works considering the mass anisotropy problem.\cite{aldrich,neet83,borges}
 In single-valley systems the conduction band  effective mass is close to
 isotropic such as the GaAs or the $\mbox{Al}_{x}\mbox{Ga}_{1-x}\mbox{As}$
 system.
 Our calculated Mott constant value for this system (i.e., $g_d=2$) is 0.23.
 Gold and Ghazali\cite{gold96} have very recently dealt with the 3D
 impurity binding energies using STLS type screening and
 numerical solution for the bound electron wave function. They reported
 for the same 3D Mott constant the value 0.25. We attribute the
 difference between our and their results to the fact that these authors
 enforced Hubbard-like form for the local-field correction which differs
 from the exact STLS local-field correction leading to a discrepancy in
 the dielectric function.

\subsection{2D EL Results}
\begin{figure}[b]
\vspace*{0cm}
\centerline{
\epsfxsize=7.5cm \epsfbox{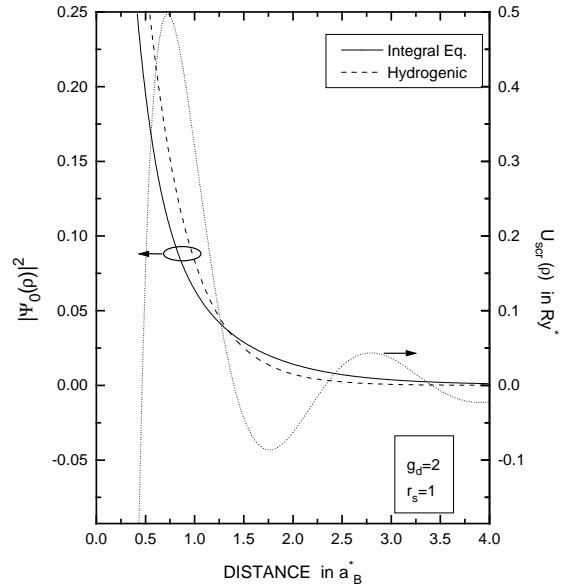}
}
\vspace*{-0cm}
\caption{Probability distribution of the bound electron wave function
 within a 2D EL having $r_s=1$ and $g_d=2$. Solid line is based on
 the integral equation solution and dashed line
 refers to 2D hydrogenic wave function, i.e., Eq.~(\protect\ref{2dwf}).
 Also shown by dotted lines is the screened potential energy
 experienced by the bound electron.}
\label{Fprobability}
\end{figure}

\begin{figure}[t]
\vspace*{-1cm}
\centerline{
\epsfxsize=7.5cm \epsfbox{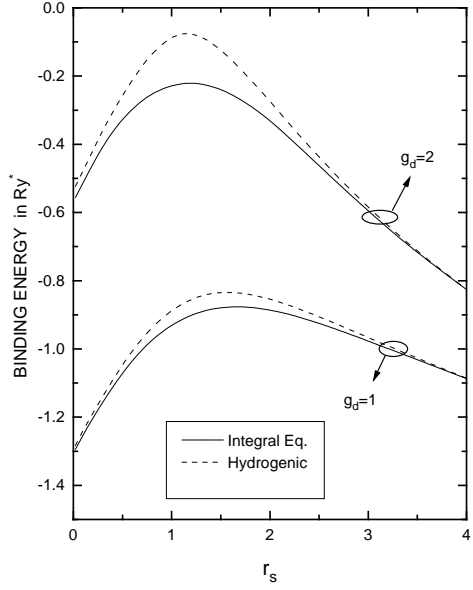}
}
\vspace*{-0cm}
\caption{Binding energy of a bound electron within a 2D EL versus
 $r_s$ for $g_d$=1 and 2. Calculations are based on the integral
 equation solution (solid lines) and 2D hydrogenic variational
 wave function (dashed lines).}
\label{Fmott2da}
\end{figure}

\begin{figure}[b]
\vspace*{0cm}
\centerline{
\epsfxsize=7.5cm \epsfbox{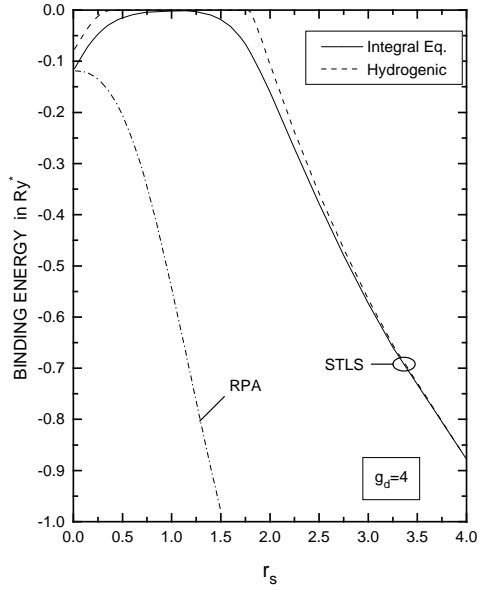}
}
\vspace*{-0cm}
\caption{Binding energy of a bound electron within a 2D EL versus
 $r_s$ for $g_d$=4 (i.e., two valley degeneracy).
 Solid line denotes the integral equation solution and the
 dashed line denotes 2D hydrogenic variational wave function
 result, both utilizing the STLS screening. Dash-dot line refers
 to RPA screening based on the integral equation solution.}
\label{Fmott2db}
\end{figure}

\begin{figure}[t]
\vspace*{-1cm}
\centerline{
\epsfxsize=7.5cm \epsfbox{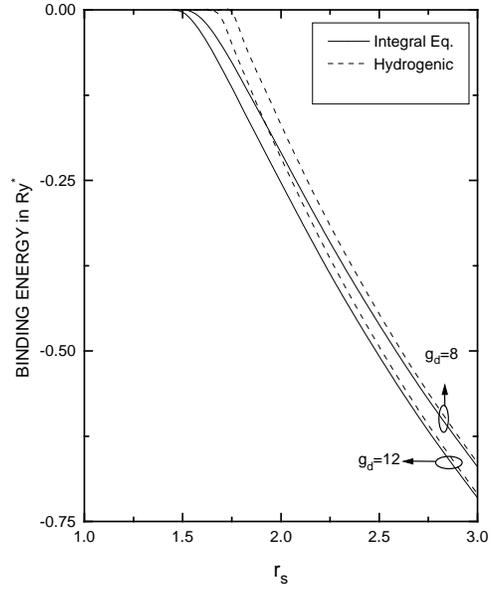}
}
\vspace*{-0cm}
\caption{Binding energy of a bound electron within a 2D EL versus
 $r_s$ for $g_d$=8 and 12 (i.e., four and six valley degeneracies).
 Solid lines denote the integral equation solutions and the
 dashed lines denote 2D hydrogenic variational wave function
 results.}
\label{Fmott2dc}
\end{figure}

\begin{figure}[b]
\vspace*{0cm}
\centerline{
\epsfxsize=7.5cm \epsfbox{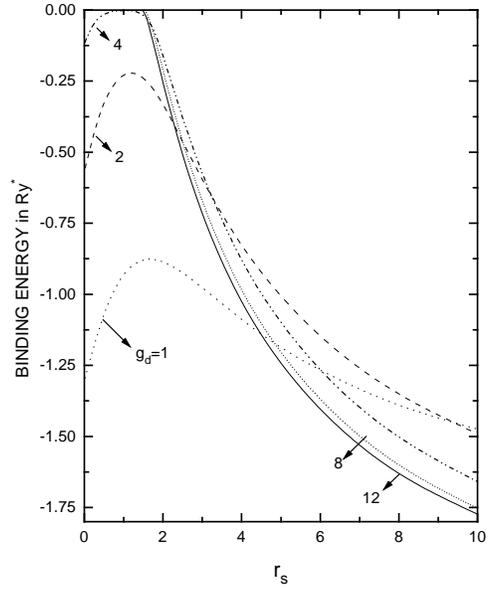}
}
\vspace*{0cm}
\caption{Binding energies in 2D EL versus $r_s$ based on the integral
 equation solutions. For comparison purposes several $g_d$ values are
 included.}
\label{Fcomb2d}
\end{figure}

In general, dimensionality is effective in almost all electronic
 properties; for our concern in 2D the role
 of Friedel oscillations is enhanced (see Fig.~\ref{Fuscr1}).
 However, quite commonly the in-plane wave function for (quasi)
 2D bound impurities\cite{panat,brum,degani,andrada} in the
 presence of free carriers
 (i.e., screening) has been chosen to be of $e^{-\rho/\lambda}$ type
 where $\lambda$ is the variational parameter. In Fig.\ \ref{Fprobability}
 the hydrogenic variational probability distribution is compared with that
 of the integral equation solution. The
 screened attractive potential energy is also added in this figure to aid
 the comparison. The probability distribution obtained by integral equation
 solution is lower in the first repulsive part of the potential energy and
 higher in the neighboring attractive region than the variational
 solution; in turn, the electron is expected to be more tightly bound.
 This is seen to be the case in  Fig.\ \ref{Fmott2da}
 showing the 2D impurity binding energy for the spin-polarized and
 normal-states respectively. Furthermore, a critical Mott density does
 not exist for these two cases and negative energy bound-states are
 available for all densities, unlike the 3D case. For $g_d=4$, the
 variational approach predicts a density range, $r_s=$ 0.38-1.81, where
 the binding energy vanishes. The integral equation solution suggests
 that this window is narrower and situated around $r_s=1$ as can be seen
 in Fig.\ \ref{Fmott2db}. As the valley-spin degeneracy further increases
 to $g_d=8$ and 12, the binding energy curves resemble those in 3D cases;
 the corresponding Mott densities are at $r_s=1.52$ and 1.48 respectively
 based on the integral equation solution whereas the variational approach
 leads to higher values (see Fig.\ \ref{Fmott2dc}). From these three
 figures we can also conclude that the hydrogenic variational technique
 is successful for the small values of the degeneracy factor, $g_d$. For
 a better comprehension, we combine all different degeneracy cases in
 Fig.\ \ref{Fcomb2d}, and present a larger density range,
 till $r_s=10$. The main observation in 2D is the remarkable dependence
 of the charged impurity binding energies on the valley-spin degeneracy
 in the presence of screening.
 
\begin{figure}[t]
\vspace*{-1cm}
\centerline{
\epsfxsize=7.5cm \epsfbox{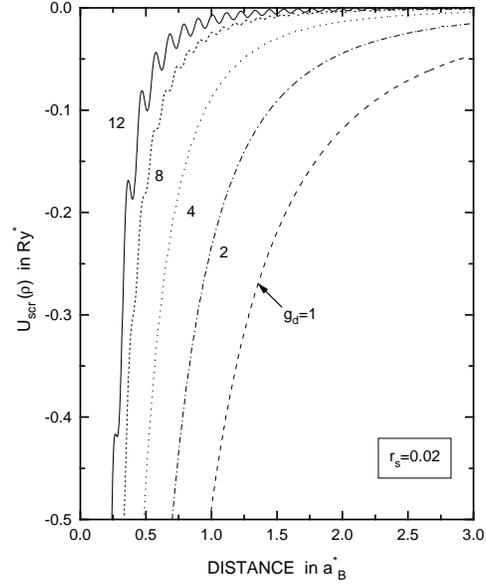}
}
\vspace*{-0cm}
\caption{Potential energy distribution due to a screened,
 singly-ionized attractive impurity versus distance. The effect of
 the degeneracy factor $g_d$ is illustrated from spin-polarized
 ($g_d=1$) to six valley degeneracy ($g_d=12$); all at a very high
 density ($r_s=0.02$) of a 2D EL.}
\label{Fuscr2}
\end{figure}

\begin{figure}[b]
\vspace*{0cm}
\centerline{
\epsfxsize=7.5cm \epsfbox{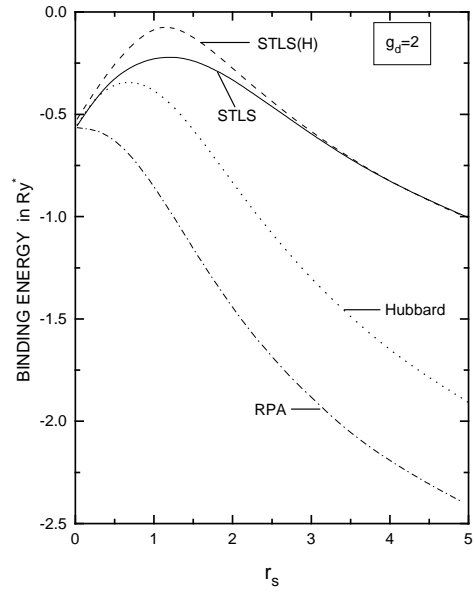}
}
\vspace*{-0cm}
\caption{The effect of dielectric function on the binding energy for
 2D EL using RPA, Hubbard and STLS screenings; all computed by
 solving the integral equation.
 Also the STLS screened binding energy is shown based the 2D
 hydrogenic variational wave function labeled by STLS(H).}
\label{Fm2diel}
\end{figure}

Fig.\ \ref{Fmott2da} shows an
 interesting strengthening of binding at the high density limit,
 $r_s \to 0$. The screened interactions for several values of $g_d$ at
 $r_s=0.02$ are plotted in Fig.~\ref{Fuscr2}. In this limit, the Friedel
 oscillations diminish and the screening is determined by the exchange
 effects. Hence, for the spin polarized case ($g_d=1$), the screening
 electrons cannot approach to the ion due to their sensible Pauli holes
 resulting in poor screening of the ion potential and enhanced binding.
 As the degeneracy parameter, $g_d$ is increased till 12, the influence
 of the Pauli hole is weakened and the central well region of the
 screened impurity potential gets narrower so that negative energy
 bound-states are no longer supported.
 Fig.\ \ref{Fm2diel} compares the effect of several dielectric functions
 (RPA, Hubbard, and STLS) on the impurity binding energy for $g_d=$2.
 Sizeable quantitative differences are observed, and STLS dielectric
 function is seen to have stronger screening power leading to a weaker
 binding. Note the agreement of the three for $r_s \to 0$ as expected.
 Hubbard dielectric function follows STLS at the high density end where
 exchange effects are
 dominant. Finally, in Fig.\ \ref{Fmott2db} we observe that for
 $g_d=4$ case RPA result becomes even qualitatively different and gives
 negative energy bound-states for all densities.

\section{Conclusion}
We investigate the role of valley-spin degeneracy on the screened
 charged impurity centers. Several complications are suppressed
 for easy comprehension and computational simplicity. These include
 the mass anisotropy, the effects of disorder and ionic lattice on the
 mobile carriers, the finite temperature, overlap of neighboring
 bound-state wave functions (impurity band formation) and the
 finite well-width in the 2D case. However, screening is treated
 using the STLS
 self-consistent local-field correction scheme and the bound electron
 wave function is handled numerically without resorting to simplistic
 approximations. We observe that care in these two points is rewarding,
 proven by the appreciable differences as compared to widely-used RPA
 and variational techniques, respectively. We anticipate that similar
 conclusions can be drawn for the Wannier excitons in the presence of 
 free
 carriers screening. Recently, Ping and Jiang\cite{ping} investigated
 the effect of screening on the exciton binding energy in
 GaAs/$\mbox{Al}_x\mbox{Ga}_{1-x}$As quantum wells using a rather simple
 approach based on the Debye screening model and a
 variational-perturbation method for the binding energy. Therefore, the
 present analysis merits to be extended to excitons.

The dependence on valley-spin degeneracy is very significant, especially
 in 2D. From the current electronic devices point of view, Si-based and
 GaAs-based devices are shown to have marked differences in the behaviour
 of screened charged impurity centers. For GaAs-based devices Pauli
 exclusion principle is more influential in the screening and impurity
 binding energies are larger than in Si-based ones. Binding energy
 dependence on the degeneracy parameter gradually saturates both in 2D
 and 3D for $g_d \ge 8$. Finally, the transport through screened charged
 impurities is also expected to have high sensitivity to the valley-spin
 degeneracy.
\acknowledgements
We are grateful to K. Leblebicio\u glu and M. Kuzuo\u glu for their
 suggestions on the operator techniques and referring us to the
 useful literature. Valuable remarks by the referee, especially on the
 device implications of our work are acknowledged.

%
%
\vspace*{4cm}
\begin{table}
\caption{Critical Mott density, $r_{sc}$ of the 3D EL. The
 corresponding Mott constants, defined as $a_{B}^{*}n_{3D}^{1/3}$ are
 indicated in parentheses. The numerical results based on the integral
 equation solution are more reliable (see text).}
\label{3dmott}
\begin{tabular}{c c c c c c}
  Degeneracy factor:  $g_d$ & 1 & 2 & 4 & 8 & 12 \\
 & {\it (spin polarized)}
 & {\it (single-valley)} & {\it (2-valley)} & {\it (4-valley)} 
 & {\it (6-valley)} \\
 \hline
 Variational-Hydrogenic: $r_{sc}$ (Mott Const.) & 2.25 (0.275) & 3.57
 (0.174) & 4.11 (0.151) & 4.13 (0.150) & 4.05 (0.153) \\
 Numerical: $r_{sc}$ (Mott Const.) &1.44 (0.430) &2.70 (0.230) &3.62
 (0.171) &3.80 (0.163) & 3.75 (0.166)\\
\end{tabular}
\end{table}
%
%
\pagestyle{empty}
\newpage
\newpage
\newpage
\newpage
\newpage

\begin{references}
\bibitem{chemical}The chemical identity of the impurity does
 not play a role in our treatment.
\bibitem{note2}Our rather simple consideration treats valley and spin
 on equal footing and their presence gets reflected to the Pauli
 exclusion principle.
\end{references}
\end{document}